\documentclass[]{aa}
\usepackage{graphicx}
\usepackage{natbib}
\bibpunct{(}{)}{;}{a}{}{,}
\graphicspath{{figures/}}
\usepackage{hyperref}
\usepackage{algorithm,algorithmicx}

\begin{document}
%
%
\title{A new approach for modelling chromospheric evaporation in response to enhanced coronal heating: II. Non-uniform heating}
\author{
C. D. Johnston\inst{1} \and A. W. Hood\inst{1} 
\and P. J. Cargill\inst{1, 2}
\and I. De Moortel\inst{1} 
}
\institute{School of Mathematics and Statistics, University of St. Andrews, St. Andrews, Fife, KY16 9SS, UK.
\and
Space and Atmospheric Physics, The Blackett Laboratory, Imperial College, London, SW7 2BW, UK.
\\
\email{cdj3@st-andrews.ac.uk}
}

%
%

  \abstract
  {
  We proposed that the use
  of an approximate \lq jump condition\rq\ at the solar  
  transition
  region permits fast and accurate numerical solutions of the
  one dimensional hydrodynamic equations when the corona
  undergoes impulsive heating. In particular, it eliminates
  the need for the very short timesteps imposed by a highly
  resolved numerical grid. This paper presents further
  examples of the applicability of the method for cases
  of non-uniform heating,  in particular, nanoflare trains
  (uniform in space but non-uniform in time)
  and
  spatially localised impulsive heating, including
  at the loop apex and base of the transition region. In all
  cases the overall behaviour of the coronal density and
  temperature shows good agreement with a fully resolved one
  dimensional model and is significantly better than the 
  equivalent results from a 1D code run
  without using the jump 
  condition but with the same coarse grid.  
  A detailed assessment of the errors introduced by the jump
  condition is presented showing that the causes of
  discrepancy with the fully  resolved code 
  are (i) the neglect of the terms corresponding to the
  rate of
  change of total energy in the unresolved atmsophere, (ii)
  mass motions at the base of the transition region 
  and (iii) for some cases
  with footpoint heating, an over-estimation of the radiative
  losses in the transition region.
  }
  \titlerunning{A new approach for modelling chromospheric
  evaporation}
  \maketitle
  
%
%

\section{Introduction
  \label{section:introduction}}
  \indent
  By using one-dimensional (1D)
  hydrodynamic (field-aligned) models
  to
  study
  the physics of magnetically confined coronal loops
  \citep[see e.g.][for a review]{paper:Reale2014},
  we have learned a great deal
  about the temporal response of coronal loop plasma to 
  heating.
  If the corona is heated by impulsive energy releases,
  both the temperature and density increase and 
  then decline, with the time of peak temperature preceding 
  that of the peak density. 
  The changes in density are the result of an enthalpy flux 
  to and from the chromosphere, through the transition region 
  (TR), 
  to the corona.
  An upward enthalpy flux
  first 
  increases the coronal density,
  often called (chromospheric) \lq evaporation\rq\
  \citep[e.g.][]{paper:Antiochos&Sturrock1978}, then after
  the time of peak density a 
  downward enthalpy flux
  drains the loop in order to power
  the TR radiation  
  \citep{paper:Bradshaw&Cargill2010a,
  paper:Bradshaw&Cargill2010b}.  
  \\
  \indent
  However, one of the main difficulties encountered with
  1D hydrodynamic computational
  models is the need to implement a grid that 
  fully resolves the steep TR, which is also very dynamic, 
  moving in response
  to the coronal heating and cooling.
  The temperature 
  length scale  
  ($L_T =T/|dT/dz|$,
  where $T$ is the temperature and
  $z$ the spatial coordinate along the magnetic 
  field) is very small in the TR.
  $L_T$ can be
  less than 1km for a loop in 
  thermal equilibrium and even shorter in hot flaring loops,
  because of the steep 
  temperature
  dependence of thermal conduction
  and the location of the
  peak of the radiative losses
  between $10^5$ \& $10^6$ K.
  Resolving these small length scales is essential in order to 
  obtain the 
  correct coronal density in response to time dependent 
  heating
  \citep[][hereafter BC13]{paper:Bradshaw&Cargill2013}. 
  Otherwise, the 
  heat flux 
  from the corona
  \lq jumps\rq\ over an under-resolved TR to the 
  chromosphere, where the incoming
  energy is then radiated away,
  rather than going into driving the evaporation.
  BC13 showed 
  that major errors in the coronal density were likely with 
  a lack of spatial resolution.
  \\
  \indent
  Furthermore,  
  in order to achieve numerical stability,
  the thermal conduction timestep scales
  as the minimum of $L_T^2$ over the whole grid implying
  long computation times for fully resolved 1D simulations.
  The problem is more severe in
  multi-dimensional 
  magnetohydrodynamic (MHD)
  models where computational resources place significant
  constraints on the achievable resolutions.
  Therefore, there is a 
  need 
  for a simple and computationally efficient method 
  that can be employed in both 1D hydrodynamic and 
  multi-dimensional 
  MHD models to help obtain the correct 
  coronal
  response to impulsive heating events.
  \\
  \indent
  \citet[][hereafter Paper I]{paper:Johnstonetal2017}
  present a physically motivated approach to 
  deal with this problem by using an integrated form of
  the energy 
  equation that essentially treats
  the unresolved region of the 
  lower TR as a 
  discontinuity. The response of the TR to changing 
  coronal conditions is then determined through the imposition 
  of a 
  jump condition, which
  compensates for the energy lost when 
  the heat flux jumps
  over an unresolved TR by imposing a local velocity 
  correction. 
  In Paper I we demonstrate that this new approach
  obtains coronal densities comparable to fully   
  resolved 1D models (e.g. BC13) but with 
  computation 
  times
  that are between one and two 
  orders of magnitude faster, since the 
  computational timestep is not
  limited by thermal conduction in the TR.  
  \\
  \indent
  However, in Paper I we consider only the case of spatially 
  uniform
  heating. The purpose of this paper is two-fold. Firstly, it
  is important to consider how the jump condition performs 
  for
  different (spatially non-uniform) heating functions and
  initial plasma conditions in order for future
  users to have
  confidence in the model. The latter is addressed through
  consideration of a nanoflare train. The former involves
  studies of highly localised heating pulses, including a
  challenging case where the heating is located at the base of
  the TR. Secondly, it has become clear that the coronal
  plasma parameters, in particular the density, show
  systematic deviations from those in fully resolved
  simulations. A full analysis of the terms in the jump
  condition (both retained and neglected) has been undertaken
  to understand the cause of this.
  This paper is not intended as a physical
  comparison between the different heating models but is
  intended to demonstrate the wide applicability of the jump
  condition used.
  \\
  \indent
  We derive 
  the jump condition and describe the implementation of the
  method in Sect.
  \ref{section:Numerical method and experiments}.
  In Sect. 
  \ref{section:Nanoflare trains} \&
  \ref{section:Non-uniform heating}, 
  we demonstrate the application of
  our approach through a series
  of examples.
  A detailed discussion of the quantities associated with
  the jump condition is presented in Sect. 
  \ref{section:Discussion_of_the_jump_condition_quantities}.
  We present our conclusions in  
  Sect. \ref{section:Conclusions}.
  
%
%

\section{Numerical method and experiments
  \label{section:Numerical method and experiments}}
\subsection{Numerical method}
  \indent
  The full details of the numerical method are 
  discussed in Paper I and so they 
  are just restated briefly here.
  We solved the 1D field-aligned MHD equations
  using 
  two different methods,
  a Lagrangian remap (Lare) approach, 
  as described for 3D MHD in
  \cite{paper:Arber2001}, adapted for 1D field-aligned
  hydrodynamics (Paper I)
  and the adaptive mesh refinement code HYDRAD
  \citep[][BC13]{paper:Bradshaw&Mason2003,
  paper:Bradshaw&Cargill2006}
  \\
  \indent
  For the initial condition we used a magnetic strand (loop)
  in static equilibrium
  of total length $2L$ that includes a 5Mm model
  chromosphere (acting mainly as a mass
  reservoir) at the base of each TR. 
  The corresponding
  temperature and density profiles were 
  derived using 
  the same base quantities
  ($T=10^4$K and n=$10^{17}$m$^{-3}$)
  and method that was described in Paper I.
  
%
%
\subsection{Overview of the unresolved transition region jump condition
  \label{section:Unresolved Transition Region Jump Condition}}
  \indent
  The 1D energy conservation equation is given by
  \begin{align}
    \frac{\partial E }{\partial t} 
    =-
    \frac{\partial}
    {\partial z} (Ev + Pv + F_c) + Q(z,t) - n^2 \Lambda(T),
    \label{eqn:1d_tec}
  \end{align}
  where
  $E = P/(\gamma - 1) + 1/2\rho v^2$ 
  (gravity is neglected for this discussion but is 
  included in the 1D field-aligned MHD equations that we 
  solve),
  $v$ is the velocity,
  $P$ is the gas pressure, $\rho$ is the density, 
  $n$ is the number density,
  $F_c=-\kappa_0 T^{5/2}\partial T/\partial z$
  is the heat flux and
  $Q(z,t)$ is a heating function that includes
  background uniform heating
  and a time dependent component that can be dependent
  on position.
  $\Lambda(T)$ is the 
  radiative loss function in an optically thin plasma, 
  which we approximate using the 
  piecewise continuous function defined in
  \cite{paper:Klimchuketal2008}.
  \\  
  \indent
  We define the unresolved transition region (UTR)
  as the region of thickness $\ell$ that extends from
  the final location in the TR at which 
  the temperature length 
  scale ($L_T$) is resolved ($z_0$, i.e.  where the criteria 
  $L_R/L_T \leq \delta < 1$ is satisfied, 
  with $L_R$ denoting the simulation
  resolution and $\delta=1/4$ is used
  throughout this paper)  
  downwards to the base of the TR ($z_b$, which is the 
  location 
  where the temperature first reaches or 
  falls below the chromospheric temperature of $10^4$K),
  as outlined in Paper I. 
  Integrating Eq.
  \eqref{eqn:1d_tec} over the UTR,
  from the base of the TR ($z_b$) upwards to the 
  top of the UTR ($z_0$), we obtain, 
  \begin{align}
    N 
    =&-
    (E_0v_0
    +P_0v_0
    +F_{c,0}
    )
    + \ell \bar{Q} - \mathcal{R}_{utr},
    \label{eqn:1d_si_utr_tec}
  \end{align}
  where $N$ is defined as
  $N\equiv\ell d\bar{E}/dt - (E_bv_b+P_bv_b+F_{c,b}) $
  and the 
  subscripts 0 and b indicate  quantities evaluated at the
  top and base of the UTR respectively. The overbars 
  indicate spatial averages over the UTR,
  $\mathcal{R}_{utr}$
  is the integrated radiative losses and 
  $\ell \bar{Q}$  is the
  volumetric heating rate
  in the UTR.
  \\
  \indent 
  We assume that the left-hand side (LHS) of Eq. 
  \eqref{eqn:1d_si_utr_tec} can be neglected
  based on the arguments presented in Paper I
  \citep[see also, e.g.][]
  {paper:Klimchuketal2008,paper:Cargilletal2012a,
  paper:Cargilletal2012b},
  so that
  $N$ represents the neglected terms.
  Hence,
  we obtain the UTR jump 
  condition,  
  \begin{align}
    \frac{\gamma}{\gamma - 1} P_0 v_0 + \frac{1}{2} \rho_0 
    v_0^3
    =
    -F_{c,0}
    + \ell \bar{Q}
    - \mathcal{R}_{utr},
    \label{eqn:1d_utr_jc}
  \end{align}
  where
  the terms on the LHS are the 
  enthalpy ($F_{e,0}$) and kinetic energy ($F_{ke,0}$) fluxes, 
  respectively. The terms on the 
  RHS are the heat flux, the integrated
  volumetric heating rate
  and the
  radiative losses ($\mathcal{R}_{utr}
  = \int^{z_0}_{z_b} n^2 \Lambda(T) \, dz$) in the UTR 
  respectively. 
  \\
  \indent  
  The lower TR is modelled as a discontinuity that responds 
  to 
  changing coronal 
  conditions through the imposition of the jump condition
  \eqref{eqn:1d_utr_jc} which in turn implies a local velocity
  correction ($v_0$), as discussed in Paper I.
  We reiterate here that to calculate this velocity 
  correction based on the equilibrium results, 
  the integrated radiative losses (IRL) in the UTR $
  (\mathcal{R}_{utr})$ are
  approximated using the radiative loss integral from the 
  resolved
  part of the upper atmosphere $(\mathcal{R}_{trc}
  = \int^{z_{{a}}}_{z_0} n^2 \Lambda(T) \, dz$
  where $z_a$ represents the loop apex),   
  that is $\mathcal{R}_{utr}=\mathcal{R}_{trc}$.
  This approximation of $\mathcal{R}_{utr}$ 
  is used only in the calculation of $v_0$.
  It remains necessary
  to
  solve 
  the full set of
  equations in the UTR (see Appendix A, Paper I) 
  in order to retain
  the structure of the TR.
  Moreover, 
  to accommodate 
  the case of 
  spatially non-uniform 
  heating, the integrated volumetric heating rate 
  ($\ell \bar{Q}$)
  is now
  calculated as, 
  \begin{align}
    \ell \bar{Q}(t) = \int_{z_b}^{z_0}
    Q(z,t) \, dz,
    \label{eqn:1d_lQ}
  \end{align}
  which reduces to $\ell Q$ for uniform heating
  (as used in Paper I).
  This new formalism \eqref{eqn:1d_lQ}
  can also be used to incorporate other
  energy deposition methods such as that from electron or ion 
  beams 
  \citep[e.g.][]{paper:Reepetal2013b}.
  \\
  \indent
  Using HYDRAD results,
  a detailed comparison of the magnitude of all the terms that
  appear in Eq. \eqref{eqn:1d_si_utr_tec} shows that
  in the neglected term $N$, $\ell d\bar{E}/dt$
  and the mass motions at the base of the TR 
  ($F_{e,b}+F_{ke,b}$)
  can have a measurable
  impact on the coronal plasma response. 
  For uniform heating this led to larger coronal densities
  than HYDRAD.
  One objective of this paper is to quantify the role of these
  neglected terms further, and this is carried out in Sect.
  \ref{section:Discussion_of_the_jump_condition_quantities}.  
  We note here that the difficulty of including the  
  $\ell d\bar{E}/dt$ term  
  remains as
  discussed in Paper I.
  In particular, if we could
  calculate $\ell d\bar{E}/dt$
  and the base motion terms accurately,
  with coarse spatial resolutions, then it would not be 
  necessary to implement the jump condition.

%
%
\subsection{Experiments
  \label{section:Experiments}}
  \indent
  To appreciate the usefulness of the UTR jump
  condition method, 
  we assess its performance
  for a much wider range of examples
  than those presented in Paper I.
  Specifically, the experiments considered 
  in this work explore nanoflare 
  trains and spatially non-uniform heating events. 
  The examples studied cover spatially
  symmetric and asymmetric 
  heating for both short and long loops.
  \\
  \indent 
  For each experiment, we assess the performance of the UTR 
  jump 
  condition method (referred to as LareJ)
  by comparing the results 
  with Lare1D (1D code run without using the jump condition but
  with the same grid)
  and HYDRAD (fully resolved 1D model). 
  LareJ and Lare1D use 
  500 uniformly spaced grid points.
  This choice of the number of grid points is  
  motivated by 
  what 
  is routinely
  used in current multi-dimensional MHD models
  \citep{paper:Bourdinetal2013,
  paper:Hansteenetal2015,
  paper:Hoodetal2016,
  paper:Dahlburgetal2016,
  paper:OHara&DeMoortel2016}, so that the simulations run in a 
  realistic time.
  \\
  \indent 
  The HYDRAD code is run in single fluid mode with a
  largest grid cell width of $400$km and 12 levels of 
  refinement employed, so that in the most highly 
  resolved regions the grid cells are of width $98$m (BC13).
  In this paper we use HYDRAD as a benchmark solution.

%
%
\section{Uniform heating: nanoflare trains
  \label{section:Nanoflare trains}}
\begin{figure*}
    \centering
    \includegraphics[width=16.5cm]
    {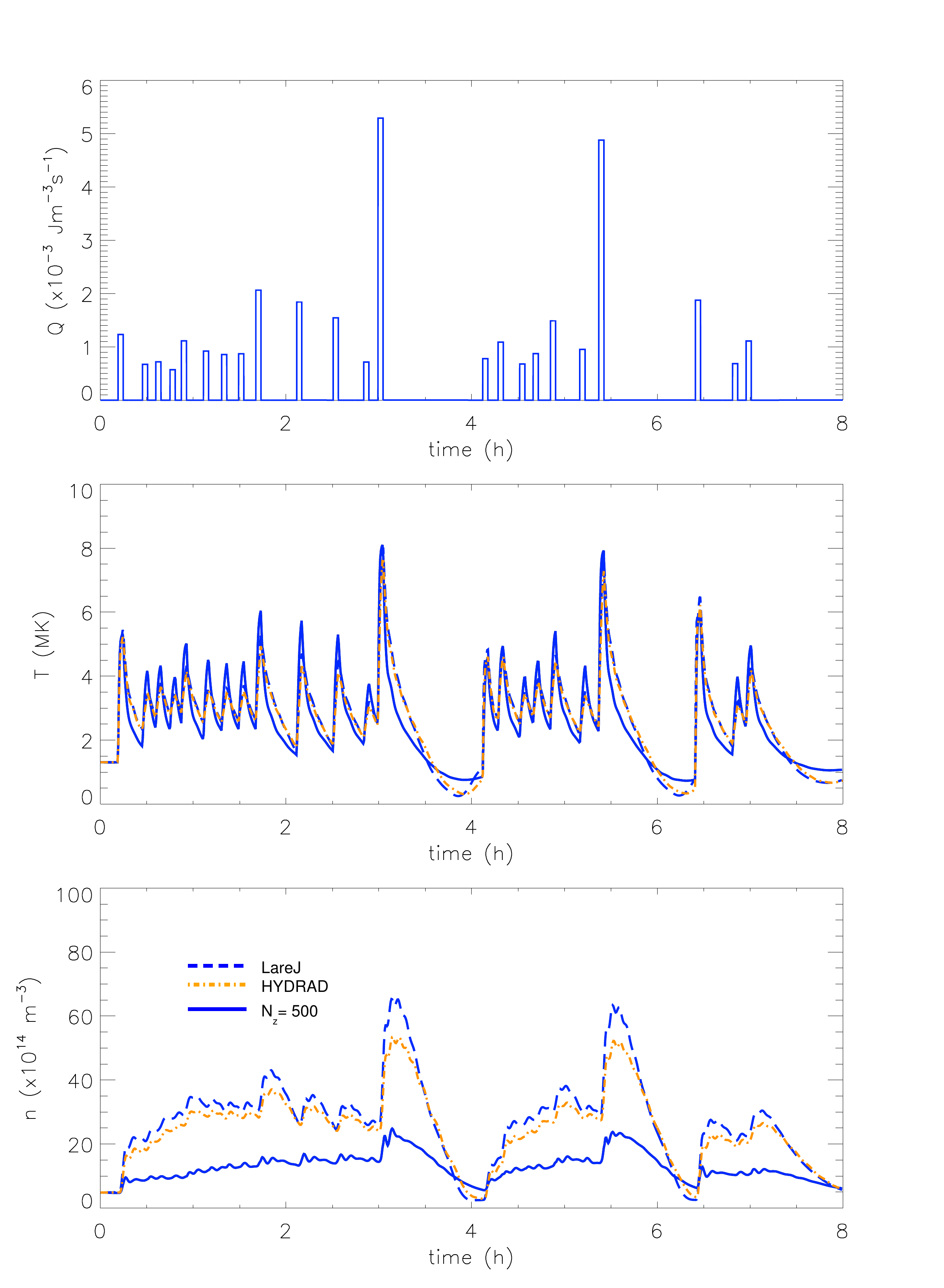}
    \caption{
    \label{Fig:nft_90Mm_loop_QTn}
    Coronal response of the plasma to a nanoflare train
    in a loop of total length 90Mm over an 8h 
    period. 
    The duration of each nanoflare is 200s.
    The panels show the volumetric heating rate and the 
    coronal averaged
    temperature and density as functions
    of time.
    The dashed blue line is the LareJ solution
    (computed with 500 grid points along the length of 
    the loop with the jump condition employed), 
    the
    solid blue line
    is the Lare1D 
    solution 
    (computed with the same spatial resolution as the
    LareJ solution but without the jump condition employed) 
    and the dot-dashed orange line 
    corresponds to the fully resolved HYDRAD solution.
    }      
  \end{figure*}

  \indent
  We consider first the case of a nanoflare train
  \citep[a sequence of heating events or nanoflares
  e.g.][]{paper:Reepetal2013a,
  paper:Cargilletal2015,
  paper:Bradshaw&Viall2016,
  paper:Barnesetal2016}.
  The aim is to investigate how the
  jump condition copes with heating events in loops that
  start with a range of 
  densities and temperatures because they
  have not undergone a full evaporation and draining cycle.
  We model a
  coronal loop of total length 90Mm and use the same
  nanoflare train as
  in \cite{paper:Cargilletal2015},
  consisting of 23 
  square wave heating pulses over an 8 hour period,
  as shown in the upper panel of
  Fig. \ref{Fig:nft_90Mm_loop_QTn}.
  This set up is  
  representative of the modelling challenge 
  faced when trying to
  understand the heating of the core loops 
  found in active regions
  \citep[e.g.][]{paper:Warrenetal2011,paper:Warrenetal2012}.
  Each nanoflare lasts 200s
  and they cover a range of magnitudes of heating,
  each 
  with an energy dependent
  waiting time between events.
  The spatial profile of the 
  heating is uniform along the loop.
  \\
  \indent
  Fig. \ref{Fig:nft_90Mm_loop_QTn}
  shows
  the coronal response 
  of the
  loop to the nanoflare train together with
  the volumetric heating 
  rate as a function of time (upper panel). 
  The central and lower panels
  show the temporal evolution of 
  the coronal averaged temperature ($T$) and density ($n$), 
  respectively. 
  The coronal averages are calculated by spatially averaging 
  over the uppermost 50\% of the loop. 
  The details of the plasma
  evolution are described in \cite{paper:Cargilletal2015}. 
  Here we focus just on the comparison between the LareJ, 
  Lare1D
  and HYDRAD solutions.
  \\
  \indent
  Comparing first LareJ and HYDRAD, we see that their
  densities follow the same basic 
  evolution, with the LareJ density larger only by about
  15--20\%, due to the neglect of  
  $\ell d\bar{E}/dt$ and the base motion terms 
  as discussed 
  above.
  The
  temperature evolution also shows good agreement.
  On the other hand,
  the Lare1D
  density is systematically lower than HYDRAD
  throughout the nanoflare train,
  on average by around a factor of 2 during small heating
  events or periods of no heating and a factor of 3 during
  larger events.
  Often there is a premature density
  peak and no substantial draining
  phase. Moreover, the density 
  being lower has two important consequences for the 
  Lare1D temperature evolution:
  (1) the peak temperature is higher
  during each heating event and
  (2)
  the subsequent cooling is more rapid.
  The first of these happens because, when
  releasing the same total amount of energy, a
  lower coronal density is heated 
  to a higher temperature.
  The second arises because
  thermal conduction is  
  more effective at lower
  density and with the temperature being higher,
  the conductive cooling timescale
  which scales as $n/T^{5/2}$ is shorter.
  This is a generic feature of the Lare1D solution.
  \\
  \indent
  Fig. 
  \ref{Fig:nft_90Mm_loop_QTn} thus shows that 
  the application of the
  jump condition approach
  is not limited to a single heating 
  and cooling cycle.  
  The approach
  deals equally well with periods of both low and high 
  nanoflare frequency.
  (Low (high) frequency nanoflares are those where the 
  waiting time between events is longer (significantly 
  shorter) than a characteristic cooling timescale.)
  This is because the underlying physics that drives the
  evaporation   
  \citep[e.g.][]
  {paper:Klimchuketal2008}
  is the same in each regime, for  
  loops that have
  cooled and drained to sub-million degree temperatures and 
  low densities between heating events
  (low frequency) and those that have
  not (high frequency).
  \\
  \indent
  We have also tested nanoflare trains in loops of total 
  length
  60Mm (short) and 180Mm (long).
  These simulations show the same fundamental 
  properties as those discussed for the 90Mm loop.
  %
  %
%
%
\section{Non-uniform heating
  \label{section:Non-uniform heating}}
  \begin{figure}
    \centering
    \resizebox{\hsize}{!}
    {\includegraphics
    {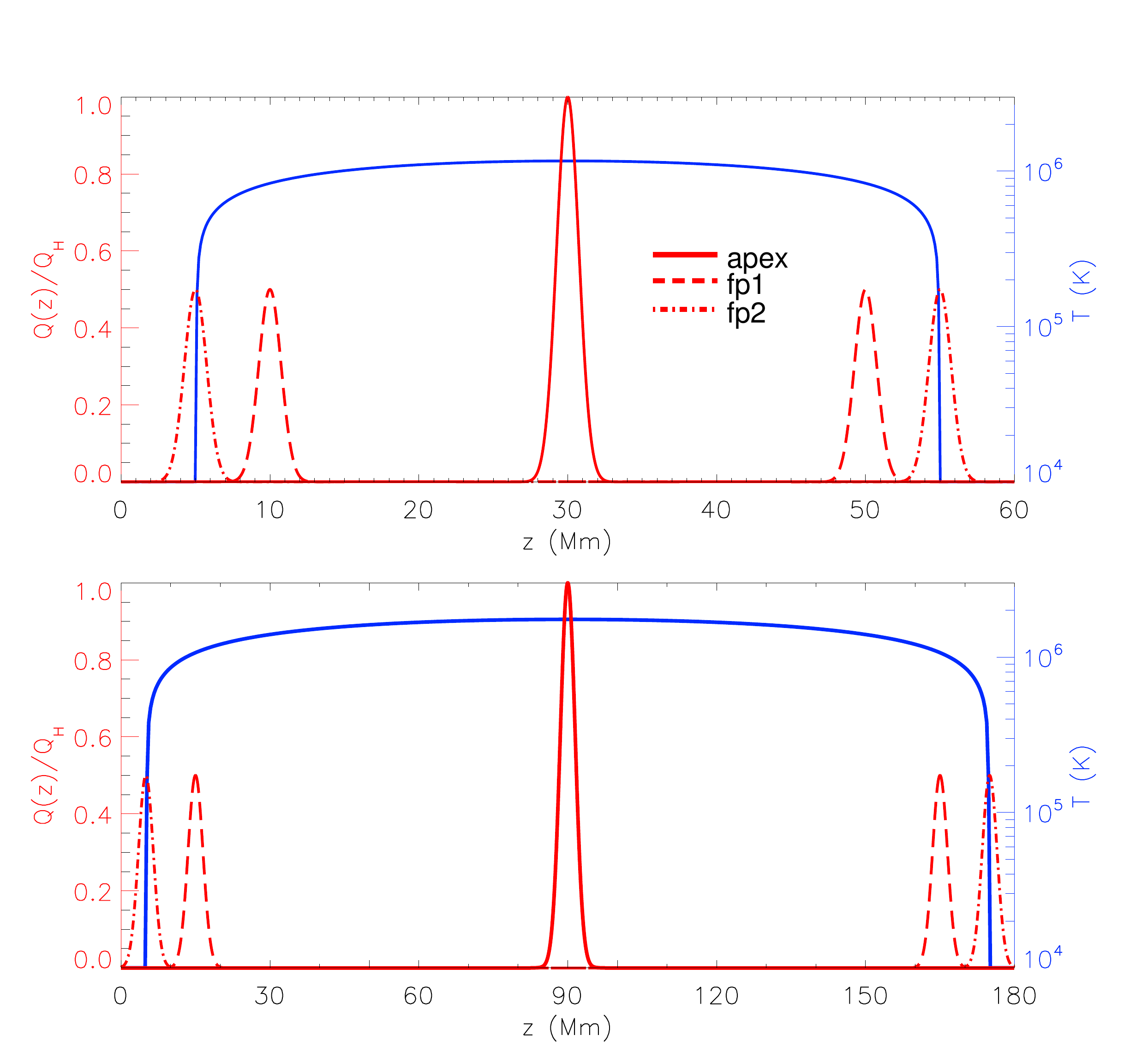}}
    \caption{
    \label{Fig:sym_sdh_profiles}
    Symmetric non-uniform heating profiles
    $Q(z)/Q_H$ (left-hand axis),
    for apex (solid red line), fp1 (base of corona,
    dashed red line) and fp2 (base of TR,
    dot-dashed red line) 
    heating (see Sect. \ref{section:Non-uniform heating}), 
    imposed on top of the temperature initial condition
    (solid blue line, right-hand axis). 
    The upper (lower) panel corresponds
    to a $60$Mm ($180$Mm) 
    loop for which we take $z_H=0.75$Mm ($z_H=1.5$Mm)
    as
    the length scale
    of heat deposition.
    }
  \end{figure}
  \begin{table*}
    \caption{
    \label{table:non_uniform_simulations}
    A summary of the parameter space used and results from
    the non-uniform heating 
    simulations.} 
    \centering
    \resizebox{\hsize}{!}
    {
    \begin{tabular}{lccccccccc}
    \hline\hline
    \\
    Case & 2L & $Q_H$ & $\tau_H$ &
    $T_{\textrm{max}}(\textrm{HYDRAD})$ &
    $T_{\textrm{max}}(\textrm{LareJ})$ &
    $T_{\textrm{max}}(\textrm{Lare1D})$ &
    $n_{\textrm{max}}(\textrm{HYDRAD})$ &
    $n(\textrm{LareJ})$ &
    $n(\textrm{Lare1D})$ 
    \\
    (Heating location)
    & (Mm) & (Jm$^{-3}$s$^{-1}$) & (s) & (MK) & 
    (MK) & (MK) & ($10^{15}$m$^{-3}$) & ($10^{15}$m$^{-3}$) &
    ($10^{15}$m$^{-3}$)
    \\
    \hline
    1 (apex) & 60  & 
    2.1   & 60  & 14.6 & 15.7
       & 15.9 & 8.8            &11.2  & 3.6
    \\
    1 (fp1)  & 60  & 
    2.1   & 60  & 10.2 & 10.9  
       &10.6 & 8.8             & 11.3 & 3.6
    \\
    1 (afp1)  & 60  & 
    2.1   & 60  & 8.0 & 8.2  
       & 8.4 & 5.4              & 7.2 & 2.4
    \\
    1 (fp2)  & 60  & 
    2.1   & 60  & 8.4  &  8.2
       & 7.6 & 8.3              &10.0 &3.3
    \\
    1 (afp2)  & 60  & 
    2.1   & 60  & 4.6  &  4.7
       & 4.6 & 4.7              & 6.5 & 2.1
    \\
    \hline
    2 (apex) & 60  & 
    $2.1\times10^{-1}$ & 600 & 7.6 & 7.7
       & 7.7 & 8.6              &10.9 & 2.7
    \\
    2 (fp1)  & 60  & 
    $2.1\times10^{-1}$ & 600 & 6.1 & 6.2
       & 5.8 & 8.9              &11.3 & 2.9
    \\
    2 (afp1)  & 60  & 
    $2.1\times10^{-1}$ & 600 & 5.4 & 5.5
       & 5.0 & 5.6              &7.1 & 2.1
    \\
    2 (fp2)  & 60  & 
    $2.1\times10^{-1}$ & 600 & 4.6 & 3.1
       & 3.8 & 9.9              &11.4 & 3.2
    \\
    2 (afp2)  & 60  & 
    $2.1\times10^{-1}$ & 600 & 4.1 & 2.2
       & 3.3 & 5.7              &5.0 & 2.4
    \\
    \hline
    3 (apex) & 180 & 
    $2.3\times10^{-1}$ & 60   &11.1 & 12.0
       & 12.0 & 0.98            &1.01 & 0.31
    \\
    3 (fp1)   & 180 & 
    $2.3\times10^{-1}$ & 60  & 5.1 & 6.5
       & 6.4  & 1.01            &1.04 & 0.32
    \\
    3 (afp1)   & 180 & 
    $2.3\times10^{-1}$ & 60  & 3.6 & 4.4
       & 4.2  & 0.63           &0.69 & 0.30
    \\
    3 (fp2)  & 180 & 
    $2.3\times10^{-1}$ & 60  & 3.7  & 3.3
       & 2.2 & 0.91             &0.81 &0.40
    \\
    3 (afp2)  & 180 & 
    $2.3\times10^{-1}$ & 60  & 2.6  & 2.5
       & 2.0 & 0.59             &0.53 &0.34
    \\
    \hline
    4 (apex) & 180 & 
    $2.3\times10^{-2}$ & 600 & 6.6 & 6.9
       & 6.9 & 0.95             &1.16 &0.43
    \\
    4 (fp1)  & 180 & 
    $2.3\times10^{-2}$ & 600 & 4.4 & 5.0
       & 4.5 & 1.04             &1.18 &0.32
    \\
    4 (afp1)  & 180 & 
    $2.3\times10^{-2}$ & 600 & 3.1 & 3.6
       & 3.4 & 0.67             &0.78 &0.29
    \\
    4 (fp2)  & 180 & 
    $2.3\times10^{-2}$ & 600 & 2.5 & 2.3
       & 2.1 & 1.31             &1.09 & 0.46
    \\
    4 (afp2)  & 180 & 
    $2.3\times10^{-2}$ & 600 & 2.0 & 1.96
       & 1.86 & 0.73            &0.67 & 0.36
    \\
    \hline
    \end{tabular}
    }
    \tablefoot{
    From left to right
    the columns show 
    the case number and heating location,
    the total length of the loop, the maximum heating rate,
    the duration of the heating pulse, the maximum averaged
    temperature attained by
    HYDRAD
    (in single fluid mode) with the
    largest grid cell of width $400$km
    and 12 levels of refinement employed,
    LareJ (computed with 500 grid points along the length of 
    the loop with the jump condition employed) and Lare1D 
    (computed with the same spatial resolution as the
    LareJ solution but without the jump condition employed)
    and their respective coronal averaged
    densities at the time of the HYDRAD
    peak density.  
    Cases 1, 2, 3 \& 4 are the equivalent of Cases 3,
    5, 9 \& 11, respectively 
    (in terms of the total energy 
    released) that were studied
    previously in Paper I.   
    fp1 and fp2 heating
    refers to symmetric footpoint
    heating concentrated at the base of the corona and
    base of the TR 
    respectively, while afp1 and afp2 heating
    are the asymmetric footpoint heating counterparts where 
    the energy is
    released only at one loop leg.
    }
  \end{table*}
  \begin{sidewaysfigure*}
    \resizebox{\hsize}{!}
    {\includegraphics{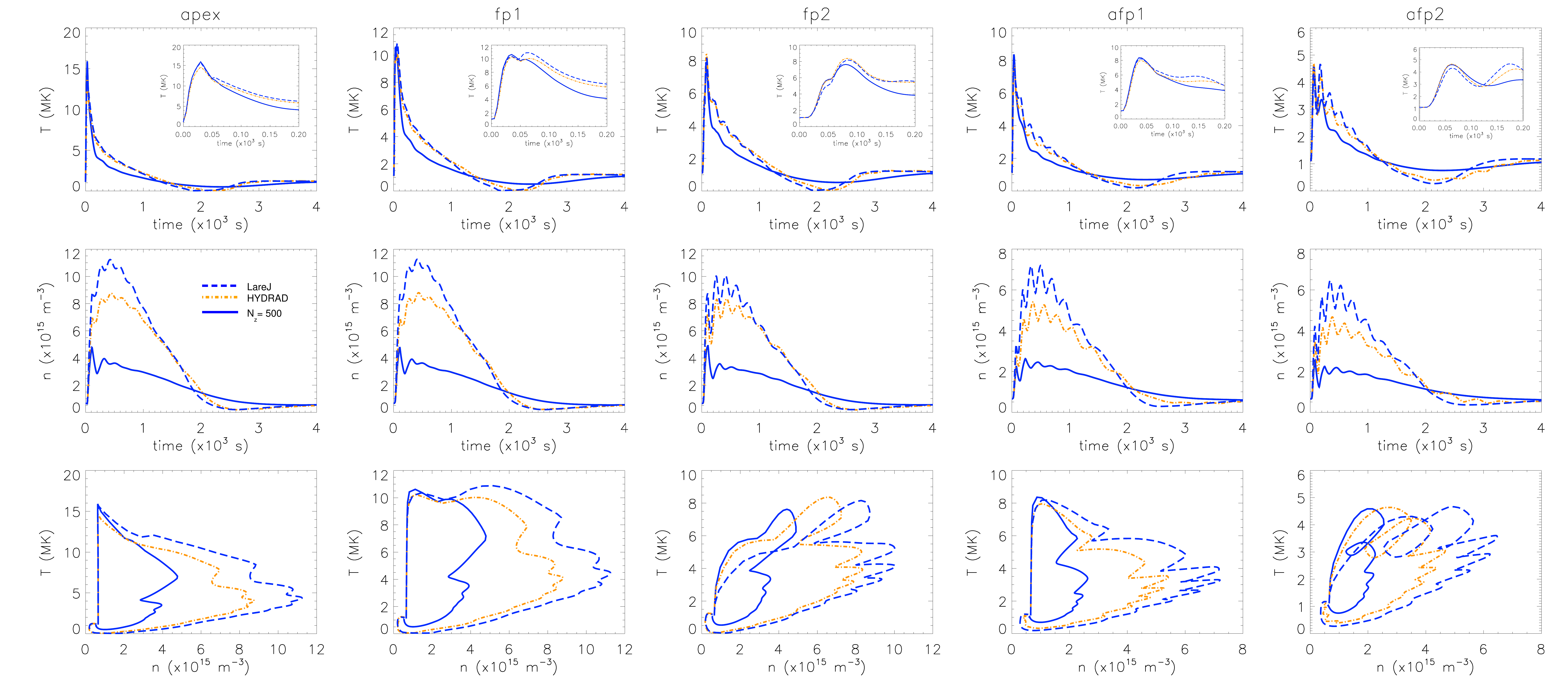}}
    \caption{
    \label{Fig:c3_sdh_ca}
    Results for Case 1. The panels show the coronal averaged
    temperature and density as functions
    of time, and the temperature versus density phase space
    plot.
    The sub-panels show the time evolution of the coronal 
    averaged temperature
    over a shorter time interval (the first 200s).
    Columns 1, 2, 3, 4 \& 5  correspond to 
    apex, fp1 
    (base of corona), fp2 (base of TR) 
    afp1 and afp2 (asymmetric)
    heating, 
    respectively.
    The dashed blue line is the LareJ solution
    (computed with 500 grid points along the length of 
    the loop with the jump condition employed), 
    the
    solid blue line
    is the Lare1D 
    solution 
    (computed with the same spatial resolution as the
    LareJ solution but without the jump condition employed) 
    and the dot-dashed orange line 
    corresponds to the fully resolved HYDRAD solution.
    The total 
    energy deposited 
    in the asymmetric heating events 
    (afp1 \& afp2)    
    is only 50\% that of the symmetric 
    heating events
    (apex, fp1 \& fp2) because only the 
    left-hand leg of the loop is heated.
    }
  \end{sidewaysfigure*}
  \begin{sidewaysfigure*}
    \resizebox{\hsize}{!}
    {\includegraphics{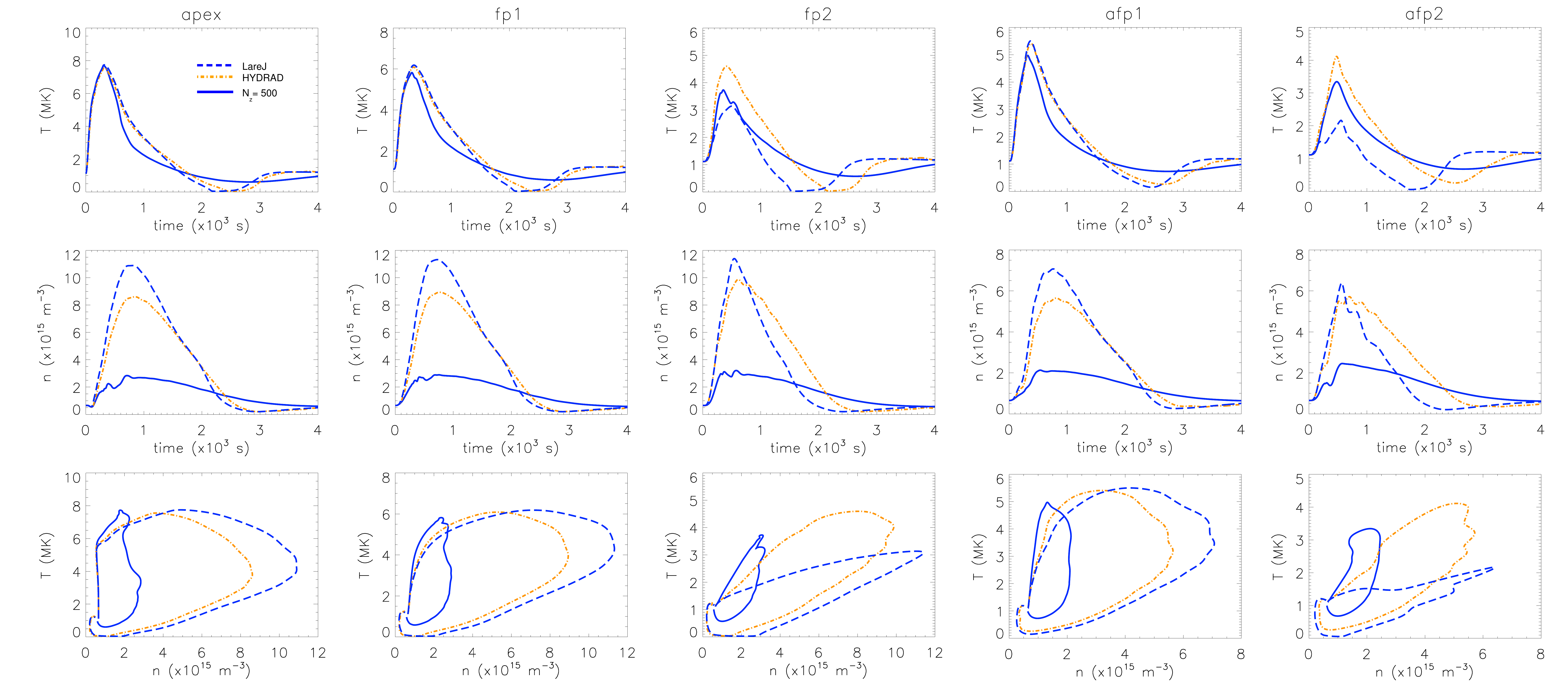}}
    \caption{
    \label{Fig:c5_sdh_ca}
    Results for Case 2.
    Notation is the same as Fig. \ref{Fig:c3_sdh_ca}.
    }
  \end{sidewaysfigure*}
  \begin{sidewaysfigure*}
    \resizebox{\hsize}{!}
    {\includegraphics{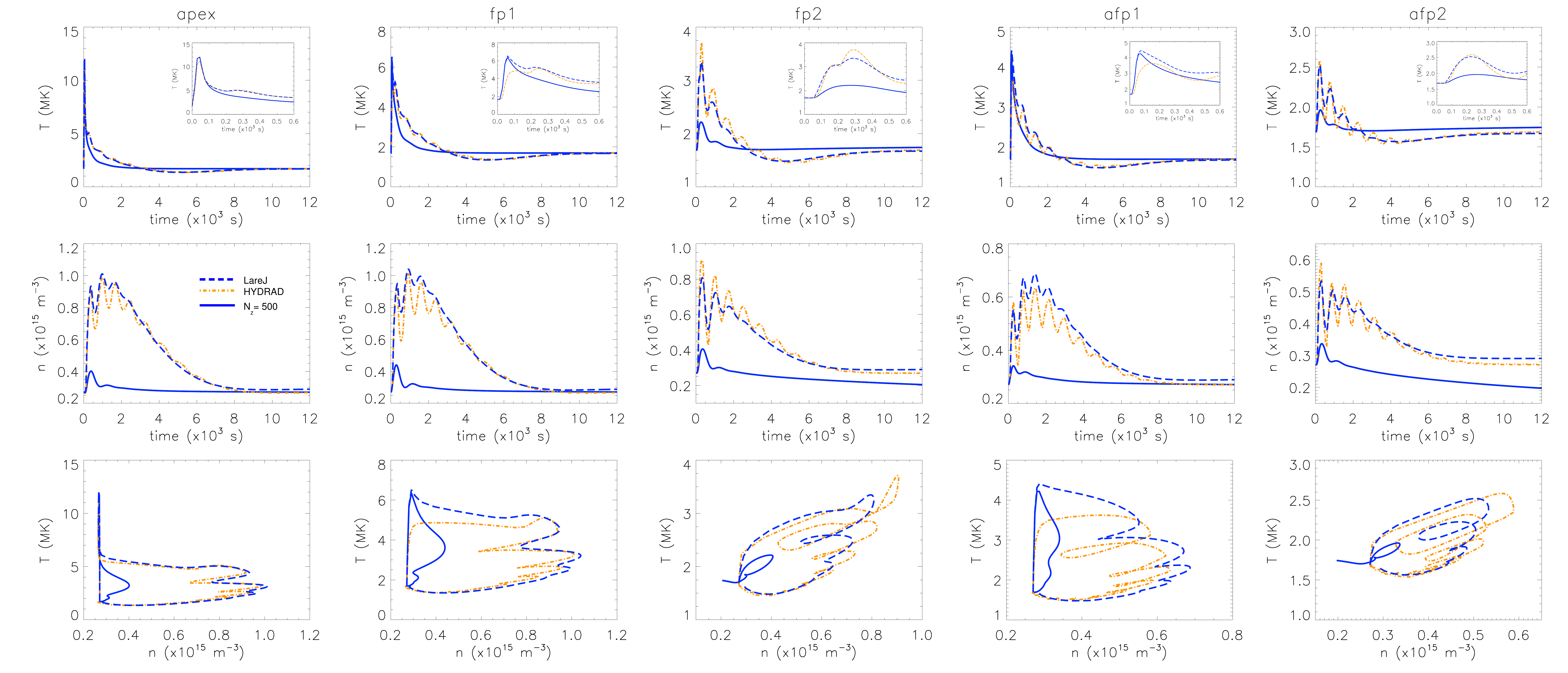}}
    \caption{
    \label{Fig:c9_sdh_ca}    
    Results for Case 3.
    Notation is the same as Fig. \ref{Fig:c3_sdh_ca} but note
    the different time axis.
    }
  \end{sidewaysfigure*}
  \begin{sidewaysfigure*}
    \resizebox{\hsize}{!}
    {\includegraphics{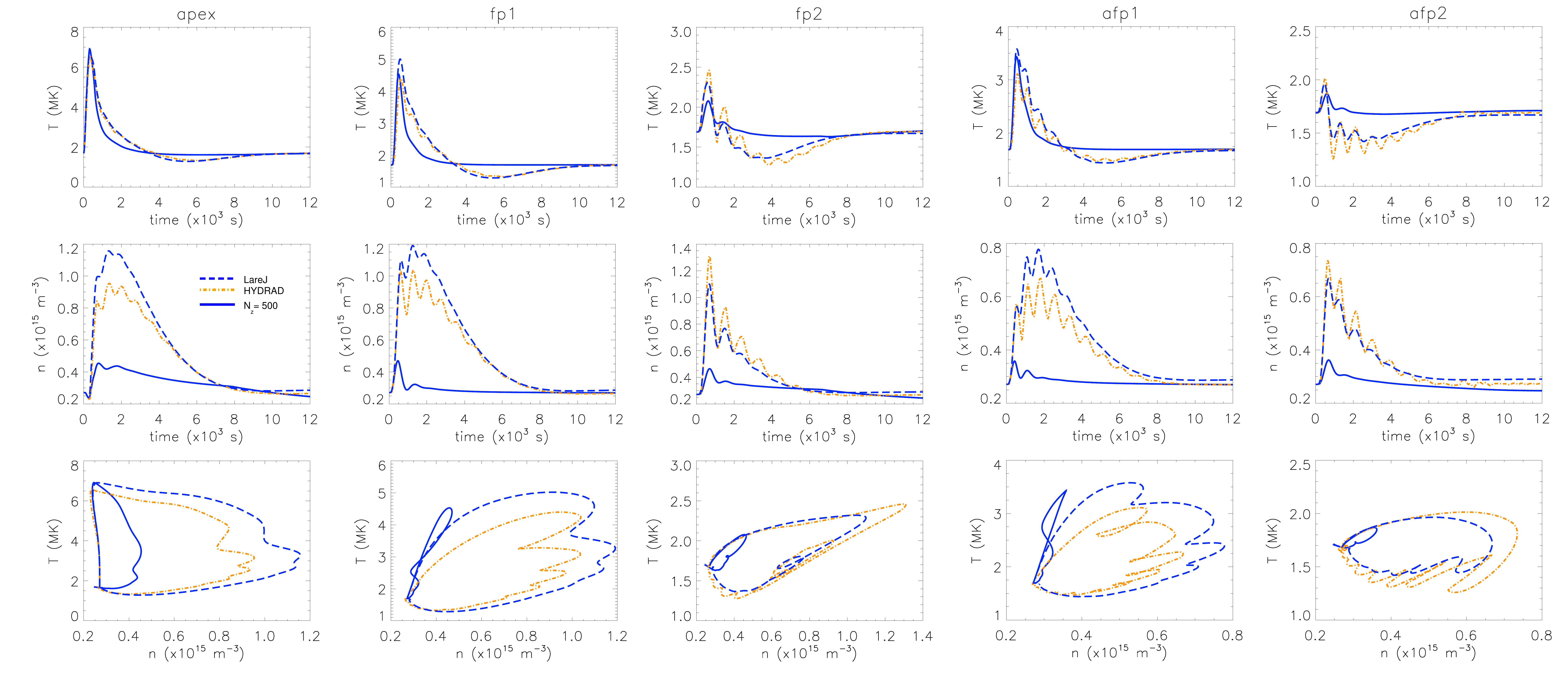}}
    \caption{
    \label{Fig:c11_sdh_ca}    
    Results for Case 4.
    Notation is the same as Fig. \ref{Fig:c3_sdh_ca} but note
    the different time axis.
    }
  \end{sidewaysfigure*}
  \indent
  Having considered the case of uniform heating 
  in Paper I and Sect. 
  \ref{section:Nanoflare trains}, we now turn our attention to
  studying spatially
  non-uniform heating
  \citep[e.g.][]{paper:Lionelloetal2009,
  paper:Mikicetal2013,
  paper:Reale2016,
  paper:Realeetal2016}. 
  In this section we  
  explore both symmetric and asymmetric heating 
  events within short (60Mm) and long (180Mm) 
  loops.
  The results are summarised in Table 
  \ref{table:non_uniform_simulations} 
  and the individual events are described below.
  \\
  \indent
  For each case we present key metrics in the table 
  that enable a comparison
  between the density and temperature attained by
  the different methods (LareJ, Lare1D and HYDRAD).
  The latter uses the maximum averaged temperature
  while the former, in contrast to Paper I, uses the
  coronal averaged density at the time of the HYDRAD peak 
  value. 
  This is a better metric for the density
  than using the maximum averaged value
  (as was done in Paper I)
  because in many cases the Lare1D density maximum
  occurs prematurely in the initial response, after
  which time the 
  accuracy of the solution fails.
  \\
  \indent
  Using the metrics it can be seen that the
  discrepancies between the LareJ and HYDRAD results
  vary from being small to significant
  while there is always significant discrepancy between
  both these methods (LareJ and HYDRAD)
  and the Lare1D density.
  Our discussion focusses primarily on the most
  difficult cases where typically the discrepancies are
  largest but it should not be ignored that the jump
  condition method in fact works very well
  (the errors are smaller) in the 
  majority of the other cases.
  \\
  \indent
  The events considered 
  are based on a subset of the cases
  that were previously studied in Paper I 
  (Cases 3, 5, 9 \& 11 there)
  now referred to as Cases 1--4 respectively.
  The energy deposition is now localised 
  to one of three distinct spatial locations. 
  These are concentrated at the loop apex and  
  the footpoints,
  as illustrated for both the short and 
  long loop
  in Fig. \ref{Fig:sym_sdh_profiles}.
  For footpoint
  heating, we consider profiles that release the 
  energy
  at the base of the corona
  (above the top of the TR) 
  and  at the base of the TR
  (i.e. $z=z_b$ in the initial equilibrium),  
  which 
  we refer to as
  fp1 and fp2 heating, respectively.
  \\
  \indent
  The temporal profile of the energy release
  is triangular, over a total duration of $\tau_H$, and
  the spatial profile is given by,
  \begin{align}
    Q(z)
    =
    Q_H 
    \exp
    \left(
    \dfrac{-(z-z_0)^2}{2z_H^2}
    \right),
    \label{eqn:sym_sdh_Q}
  \end{align}
  where $z_0$ is the location of maximum heating, $z_H$ 
  is the length scale of heat deposition and $Q_H$ is the
  maximum heating rate.
  We relate the results to Paper I by releasing the
  same total
  amount of energy in the symmetric non-uniform heating events
  as was released in the corresponding uniform heating 
  ($Q_{U}$) cases. 
  Hence, the maximum heating rate is calculated
  as,
  \begin{align}
    Q_H 
    =
    Q_{U} 2L_C/z_H\sqrt{2\pi},
    \label{eqn:sym_sdh_Q_H}
  \end{align}
  where $2L_C$ is the total coronal length
  between the TR bases.
  \\
  \indent
  Asymmetric heating is studied by adjusting
  the symmetric footpoint heating profiles 
  to release energy only at the 
  left-hand leg of the loop. Thus, the total energy deposited
  in the asymmetric heating events (referred to
  as afp1 and afp2 heating) is only 50\% 
  that of the symmetric heating counterparts,
  but the heating at the left footpoint is the same.

%
%
\subsection{Case 1
  \label{section:Case1}}
  \indent
  Case 1 is representative of a small flare in a short loop.
  For uniform heating (Paper I),
  this case proved to be the most challenging one
  for obtaining agreement between LareJ and HYDRAD. 
  Fig. \ref{Fig:c3_sdh_ca} shows
  the temporal evolution of 
  the coronal averaged temperature ($T$), density ($n$)
  and the corresponding temperature versus
  density phase space plot,
  in response to apex, 
  fp1, 
  fp2,
  afp1 and
  afp2 heating (columns 1--5).
  \\
  \indent
  Fig.
  \ref{Fig:c3_sdh_ca} demonstrates
  the level of agreement between the LareJ and HYDRAD
  solutions.
  Starting with the plasma response to apex heating,
  column 1 of Fig.
  \ref{Fig:c3_sdh_ca}
  shows that the LareJ density has the same generic 
  evolution with respect to the evaporation 
  required by the heating, the time of peak density and the 
  subsequent decay phase.
  The discrepancy between the LareJ density and the
  HYDRAD density
  is about 30\% at the density peak, 
  the source of which will be discussed in detail
  in Sect.
  \ref{section:Discussion_of_the_jump_condition_quantities}
  but we note that this
  is of the same order as  found in 
  Paper I.
  Furthermore, any differences 
  in the temperature are smaller than 
  those in the density, because of its 
  weaker dependency
  on the spatial resolution for this class of problem (BC13).
  It can also be seen that LareJ represents a considerable 
  improvement on the 
  Lare1D solution, whose temperature and density suffer from 
  rapid cooling and a premature density peak
  as noted in 
  Sect. \ref{section:Nanoflare trains}.
  \\
  \indent
  Due to the fact that thermal conduction is very efficient
  at coronal temperatures,
  the density response to fp1 heating is 
  similar to apex heating (and both are 
  similar to uniform heating). Following the energy 
  release in fp1 heating there is an upward propagating 
  conduction front that heats the coronal plasma 
  (so the peak temperature is lower than for apex heating)
  and a downward propagating front that drives the 
  evaporation from the TR.  
  Fig. \ref{Fig:c3_sdh_ca} and Table 
  \ref{table:non_uniform_simulations}
  shows that
  the agreement obtained between the LareJ and HYDRAD 
  solutions 
  (for the fp1 heating event)
  is similar to apex heating.
  The Lare1D solution 
  has the same problems as before.
  \\
  \indent
  For fp2 heating,
  the energy deposition is centred on the base of the TR and 
  so is deposited
  in the chromosphere and UTR.
  Hence, the coronal plasma is heated by an upward propagating 
  conduction front while simultaneously
  the evaporation (density front) from the TR
  is driven by a combination of the
  local energy release and
  conductive heating.
  Therefore, fp2 heating 
  poses different challenges from
  apex and fp1 heating.
  In particular a difficulty with fp2 heating is that 
  part of
  the energy released may be lost to artificially large
  radiation within the UTR (see Sect.
  \ref{section:Sources_of_error_under_evaporation}).
  \\
  \indent 
  However, column 3 in Fig. 
  \ref{Fig:c3_sdh_ca}
  shows that for the   
  fp2 heating event considered in Case 1, the agreement  
  between the temporal evolution of the LareJ temperature and 
  density and the corresponding HYDRAD quantities is 
  respectable and still significantly better than Lare1D.
  The 
  fact that
  the HYDRAD peak temperature now 
  marginally exceeds that of the LareJ 
  solution demonstrates
  that energy was lost lower down in 
  the UTR
  (usually 
  the reverse is true, see Table 
  \ref{table:non_uniform_simulations}),
  although it was only a small amount
  for this
  particular heating event.
  \\
  \indent
  The response of the TR 
  to afp1 and afp2 asymmetric footpoint heating is  
  different for each loop leg.
  On one hand the initial response at the 
  left-hand leg of the loop
  is equivalent
  to the symmetric footpoint heating events,
  while on the other hand the response at the right-hand leg
  is consistent with a weakened apex heating event 
  because it 
  only undergoes heating following the arrival of the
  conduction 
  front that is 
  launched from the left-hand leg.
  In accordance with these 
  differences in the TR response,
  Fig. 
  \ref{Fig:c3_sdh_ca} shows that 
  for the afp1 and afp2 heating events,
  the coronal temperature and 
  density peaks are lower
  than the equivalent 
  symmetric heating quantities.
  However, 
  the level of agreement between the LareJ and HYDRAD solutions
  is similar to that discussed above for the
  symmetric heating events.
  Thus, the jump condition model
  can be employed to capture
  the coronal response to  both symmetric and asymmetric
  non-uniform heating events.
  \\
  \indent
  In summary, for Case 1,
  the jump condition solutions (LareJ) obtain a coronal 
  density comparable 
  to HYDRAD (fully resolved 1D 
  model) but with a significantly faster computation time
  (the gains are consistent with those presented 
  in Paper I, the speed-up is between one and two  
  orders of magnitude)  
  and the approach
  significantly improves the accuracy of both the 
  coronal density and temperature evolution when 
  compared to the equivalent simulations run without 
  the jump condition (Lare1D, the solid blue lines).
  So despite the complexity of the type of heating 
  considered,
  the jump condition still produces acceptable results when 
  using coarse resolution.
  
%
%
\subsection{Case 2}
  \indent
  Fig. \ref{Fig:c5_sdh_ca} shows 
  the results for Case 
  2.
  Here, the total amount of energy released is the
  same as Case 1 
  but the heating is slower, taking place over a longer 
  timescale (600s).
  When compared with HYDRAD,
  the performance of the LareJ solution 
  once again shows good agreement, 
  for the apex, fp1 and 
  afp1 heating events.
  This is expected
  because the 
  terms that
  control the evaporation
  are essentially the same 
  as those in Case 
  1, but
  acting over longer timescales.
  \\
  \indent
  On the other hand, 
  for the fp2 heating event the
  LareJ 
  solution
  has its largest discrepancies
  (in response to symmetric energy deposition)
  when compared 
  over the complete heating 
  and cooling cycle with
  the fully resolved 1D model (HYDRAD). 
  Column 3 in Fig. \ref{Fig:c5_sdh_ca}
  shows that the temperature
  is lower and while the density gives a good description
  up until the time of the maximum, the draining phase 
  begins slightly early.
  These discrepancies
  are related to energy losses in the UTR with LareJ. This
  will be discussed in detail
  in Sect.
  \ref{section:Sources_of_error_under_evaporation}.
  Similar problems are also observed in the afp2 heating event
  but the energy losses are more significant for
  asymmetric
  footpoint
  heating (afp2) because there is only one upward propagating
  conduction front.
  \\
  \indent
  However, when we compare the LareJ solution with the
  equivalent simulation run without 
  the jump condition (Lare1D), 
  it remains clear that we still
  significantly improve the accuracy of the 
  coronal density. Hence, we capture a more realistic
  evolution in response to the heating
  by employing the 
  jump condition (LareJ),  
  despite the energy 
  losses in the UTR.
  
%
%
\subsection{Cases 3 \& 4}
  \indent
  We present
  the results for the long ($180$Mm) loop simulations
  in Fig. 
  \ref{Fig:c9_sdh_ca} \& \ref{Fig:c11_sdh_ca}.
  In all of the ten heating events considered in
  Cases 3 \& 4, the figures 
  show that the LareJ solutions significantly improve
  the Lare1D results to 
  accurately capture the coronal response 
  of the fully resolved 1D model (HYDRAD). 
  The details are analogous to the
  short loop simulations
  previously discussed
  with only one exception. Namely,
  in the fp2 and afp2 heating events,
  the LareJ
  density peak somewhat underestimates
  the HYDRAD value whereas in general 
  the reverse is true (see Table 
  \ref{table:non_uniform_simulations}).
  The explanation for these two different types of  
  behaviour, over and under-evaporating,
  will be discussed next
  in Sect. 
  \ref{section:Discussion_of_the_jump_condition_quantities}.
  
%
%
\section{Discussion of the quantities associated with the
  UTR jump condition
  \label{section:Discussion_of_the_jump_condition_quantities}}
  \begin{figure*}
    \centering
    \includegraphics[width=15cm]
    {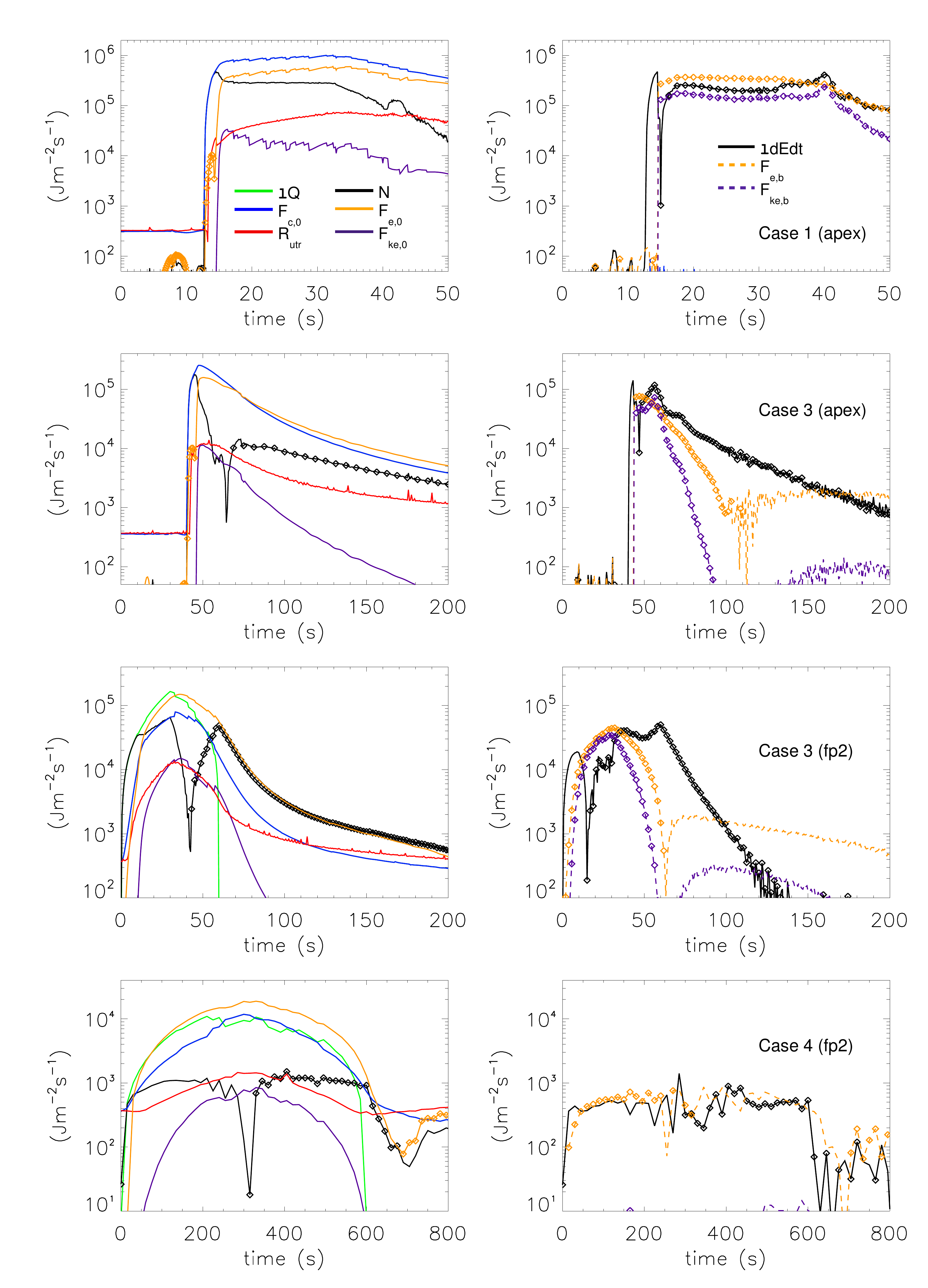}
    \caption{
    \label{Fig:sources_of_error_HYDRAD}
    Analysis of quantities associated with the UTR jump
    condition based on fully resolved HYDRAD simulations.
    Rows 1, 2, 3 \& 4 correspond to 
    Case 1 apex, Case 3 apex,
    Case 3 fp2 and Case 4 fp2 
    heating, 
    respectively.
    The left-hand panels show the terms in Eq. 
    \eqref{eqn:1d_si_utr_tec}
    that control
    the plasma response, namely
    the volumetric heating rate in the UTR ($\ell \bar{Q}$, 
    green line),
    heat flux at the top of the UTR ($F_{c,0}$, blue line),  
    IRL in the UTR ($R_{utr}$, red line), 
    neglected terms ($N$, black line,
    the LHS of Eq. \eqref{eqn:1d_si_utr_tec}),
    enthalpy flux at the top 
    of the UTR ($F_{e,0}$, orange line) and the
    kinetic energy flux at the top 
    of the UTR ($F_{ke,0}$, purple line).
    In the upper two left-hand panels (Cases 1
    \& 3
    apex heating) 
    there is no green line ($\ell \bar{Q}$)
    because there is no heating in
    the UTR except that from the background.
    The right-hand panels show the breakdown of the neglected 
    terms 
    ($N$), consisting of 
    $\ell d\bar{E}/dt$ (black line), the
    enthalpy flux at the base of the TR ($F_{e,b}$, dashed 
    orange 
    line) and the
    kinetic energy flux at the base of the TR
    ($F_{ke,b}$, dashed purple line).
    The
    heat flux at the base of the TR ($F_{c,b}$)
    is not shown in
    the breakdown of the neglected 
    terms 
    ($N$)
    because it is always negligible.
    Lines connected by diamond symbols indicate negative
    quantities and lines without 
    diamonds 
    indicate positive quantities.   
    }
  \end{figure*}
  \begin{table*}
    \caption{
    \label{table:apex_fp1_simulations_jc_terms}
    A summary of the terms from Eq. \eqref{eqn:1d_si_utr_tec}
    that control the plasma response to the apex and fp1 
    heating events, considered in Sect. 
    \ref{section:Non-uniform heating}, 
    from the HYDRAD results.} 
    \centering
    \resizebox{\hsize}{!}
    {
    \begin{tabular}{lcccc}
    \hline\hline
    \\
    Case &
    $F_{c,0}$ balances $\mathcal{R}_{utr}$ & 
    $\ell d\bar{E}/dt$ balances $F_{c,0}$  & 
    $F_{c,0}$ drives upward $F_{e,0}$ &
    First density peak 
    \\
    (Heating location) &
    &
    &
    ($F_{e,0}$ reduced due to terms associated with $N$)
    &
    (peak density)
    \\
    \hline
    1 (apex) & 0--12s  & 12--15s & 15--100s (15--40s)
    & 100s (500s)
    \\
    1 (fp1) & 0--5s  & 5--8s & 8--100s (8--35s)
    & 100s (500s)  
    \\
    \hline
    2 (apex) & 0--65s  & 65--80s & 80--500s (80--150s)
    & 800s
    \\
    2 (fp1) & 0--20s  & 20--40s & 40--500s (40--100s)
    & 800s
    \\
    \hline
    3 (apex) & 0--40s  & 40--45s & 45--270s (45--50s) 
    & 300s (1000s)
    \\
    3 (fp1) & 0--13s  & 13--25s & 25--230s (25--30s)
    & 300s (1000s)  
    \\
    \hline
    4 (apex) & 0--190s  & 190--230s & 230--1000s (230--300s) 
    & 700s 
    (1500s)
    \\
    4 (fp1) & 0--50s  & 50--90s & 90--900s (90--120s) 
    & 500s (1500s)
    \\
    \hline
    \end{tabular}
    }
    \tablefoot{
    The columns show
    the time interval during which 
    conductive heating ($F_{c,0}$) balances
    the radiative losses ($\mathcal{R}_{utr}$), the
    rate of change of total energy in the UTR
    ($\ell d\bar{E}/dt$) 
    balances the conductive heating,
    conductive losses drive an upward enthalpy flux
    (with the upflow, $F_{e,0}$, reduced due to terms
    associated with $N$) 
    and the time of the first density peak (and 
    subsequent peak density).
    }
  \end{table*}
  \begin{table*}
    \caption{
    \label{table:fp2_simulations_jc_terms}
    A summary of the terms from Eq. \eqref{eqn:1d_si_utr_tec}
    that control the plasma response to the fp2 
    heating events, considered in Sect. 
    \ref{section:Non-uniform heating}, from the HYDRAD 
    results.} 
    \centering
    \resizebox{\hsize}{!}
    {
    \begin{tabular}{lccccc}
    \hline\hline    
    \\
    Case &
    Total local heating  & 
    Imbalance drives & 
    Local heating drives upward $F_{e,0}$ &
    Upflow associated with  &
    First density peak 
    \\
    (Heating location) &
    balances total losses  &
    upward $F_{e,0}$ & 
    ($F_{e,0}$ reduced due to terms associated with $N$)
    &
    inclusion of terms in $N$ &
    (peak density)
    \\
    \hline
    1 (fp2) & 0--5s  & 5--20s & 20--60s (5--30s) & 
    60--80s & 100s (500s) 
    \\
    \hline 
    2 (fp2) & 0--10s & 10--50s & 50--450s 
    (10--100s) &
    -- & 600s
    \\
    \hline 
    3 (fp2) & 0--10s  & 10--35s & 35--60s 
    (10--35s) & 
    60-200s & 300s  
    \\
    \hline 
    4 (fp2) & 0--20s & 20--50s & 50--600s
    (20--100s) &
    -- & 700s
    \\
    \hline
    \end{tabular}
    }
    \tablefoot{
    The columns show
    the time interval during which the total local heating
    ($F_{c,0}+\ell \bar{Q}$)
    balances
    the total losses ($\ell d\bar{E}/dt + \mathcal{R}_{utr}$),
    the imbalance between these terms 
    ($F_{c,0}+\ell \bar{Q}
    -N - \mathcal{R}_{utr}$)
    drives an upward enthalpy flux ($F_{e,0}$),
    local heating drives an upward enthalpy flux 
    (with the upflow reduced due to terms
    associated with $N$), 
    the upflow is associated with the inclusion of terms in
    $N$,
    and the time of the first density peak (and 
    subsequent peak density).
    }
  \end{table*}
  \begin{figure*}
    \centering
    \includegraphics[width=18cm]
    {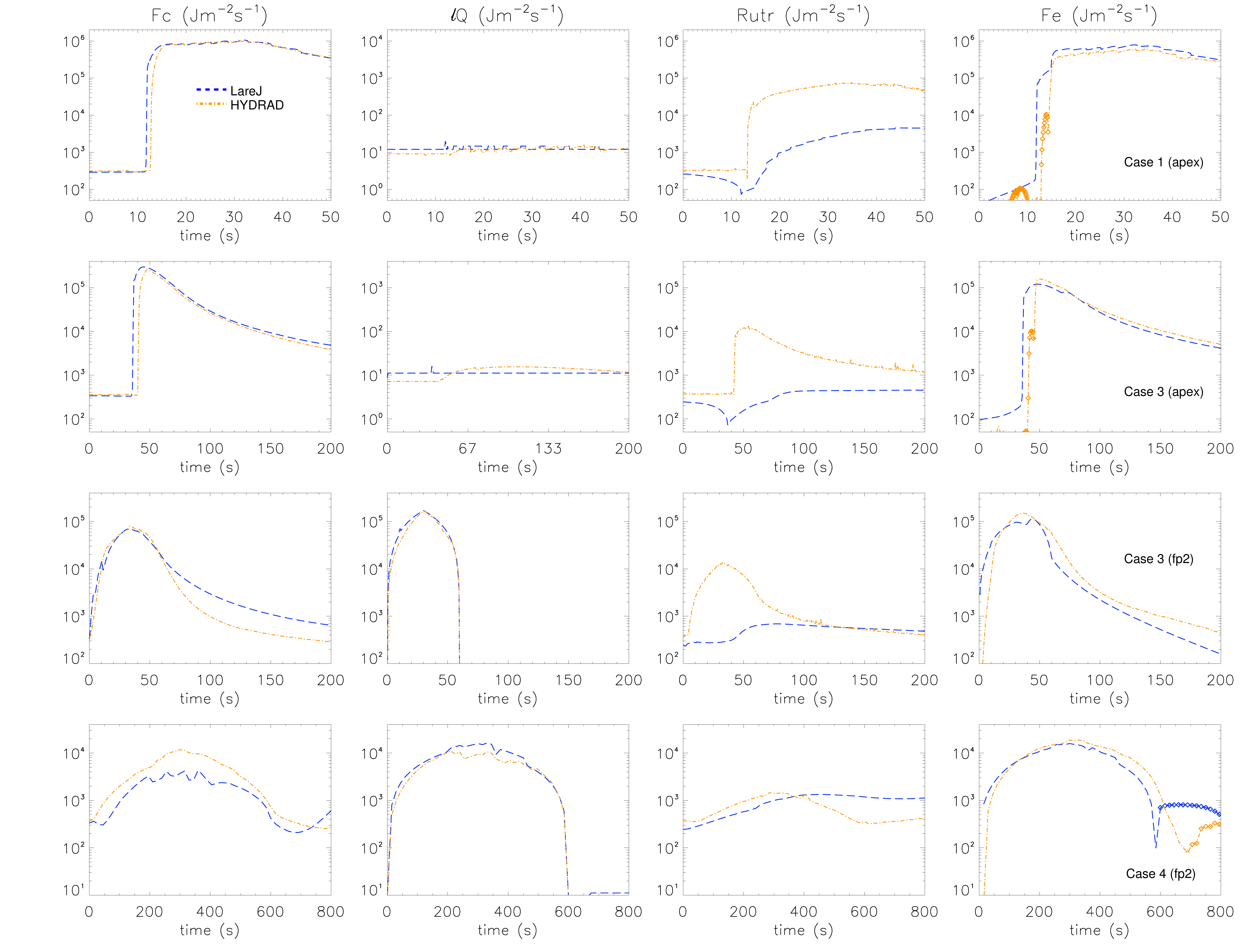}
    \caption{
    \label{Fig:sources_of_error_LareJ}
    Comparison of the dominant quantities in the 
    UTR jump
    condition obtained from the HYDRAD and LareJ 
    solutions.
    Rows 1, 2, 3 \& 4 correspond to  
    Case 1 apex, Case 3 apex,
    Case 3 fp2 and Case 4 fp2 
    heating, respectively.
    From left to right the columns show 
    the heat flux at the top of UTR ($F_{c,0}$),  
    volumetric heating rate in the UTR ($\ell \bar{Q}$),
    IRL in the UTR ($\mathcal{R}_{utr}$)
    and
    enthalpy flux at the top 
    of the UTR
    ($F_{e,0}$, lines connected by diamond symbols indicate 
    where 
    the enthalpy flux is downflowing and lines without 
    diamonds 
    indicate where the enthalpy flux is 
    upflowing).
    The upper two rows 
    (Cases 1 \& 3 apex heating) 
    in the second column ($\ell \bar{Q}$) 
    show only the background heating
    and have a different vertical axis.
    The dashed blue line is the LareJ solution
    (computed with 500 grid points along the length of 
    the loop with the jump condition employed)
    and the dot-dashed orange line 
    corresponds to the fully resolved HYDRAD solution.
    }
  \end{figure*}
  \indent
  We now turn to a more detailed discussion of the
  results
  obtained with the different methods used in each experiment.
  The principal issue to be addressed is the discrepancy
  between the density evolution in LareJ and HYDRAD. In most
  cases the LareJ peak density exceeds that obtained by HYDRAD
  (referred to as over-evaporation), although for some
  footpoint heating cases, the opposite is true 
  (under-evaporation). (The causes for the differences between
  the under-resolved Lare1D and the other simulations have been
  discussed earlier and in Paper I and will not be considered
  further.) This comparison of the models is undertaken through
  an analysis of the terms associated with the UTR jump
  condition \eqref{eqn:1d_utr_jc}, 
  in particular the various terms in the
  definition of the terms neglected in 
  LareJ ($N$: Eq. \eqref{eqn:1d_si_utr_tec}). The
  cases where the LareJ density exceeds (falls below) the
  HYDRAD value are discussed in Sect. 
  \ref{section:Sources_of_error_over_evaporation} 
  (\ref{section:Sources_of_error_under_evaporation}). 
  \\
  \indent
  In general, there are two important terms in $N$. One is the
  rate of change of total energy in the UTR 
  ($\ell d\bar{E}/dt$)
  and when
  this is positive (negative), the corrected upflow ($v_0$) in
  LareJ is enhanced (decreased) over the upflow at the same
  point in the HYDRAD solution. The second important term in 
  $N$
  involves the mass motions at the base of the TR 
  ($F_{e,b}+F_{ke,b}$). 
  When the TR moves downwards (upwards) the neglect of
  this motion in LareJ leads to enhanced (decreased) values of
  $v_0$. 
  The heat flux at the base of the TR ($F_{c,b}$) is negligible
  for all cases.
  \\
  \indent
  An analysis of the resolved HYDRAD results enables us to
  quantify the importance of the terms neglected in LareJ, and
  Fig. \ref{Fig:sources_of_error_HYDRAD}
  shows the quantities obtained by HYDRAD that are
  present in the UTR equations. The left column shows the
  various terms from Eq. \eqref{eqn:1d_si_utr_tec}
  and the right breaks down $N$ into
  its important components. 
  Fig. \ref{Fig:sources_of_error_LareJ}
  shows quantities in Eq. \eqref{eqn:1d_utr_jc}
  obtained from the HYDRAD and LareJ simulations. 
  The definition of the \lq UTR\rq\ is based on the
  time evolution of the temperature from the LareJ solution,
  though of course it is fully resolved in HYDRAD.
  Apex heating for Cases 1 \& 3 are shown in the upper two
  rows,
  only for the first 50s in Case 1 and the first 200s in Case
  3. The lower two rows show the fp2 heating events for Cases 3
  \& 4 where the LareJ density peak falls below that given by
  HYDRAD. The first 200s are shown for Case 3 in row 3 and the
  first 800s for Case 4 in row 4. We only consider the time
  evolution until the first density peak because this time
  interval is the main source of error in the subsequent peak
  density.
  
%
%
\subsection{Sources of error: over-evaporation (apex \&
  fp1 heating)
  \label{section:Sources_of_error_over_evaporation}}
  \indent
  For the Case 1 apex heating event, the HYDRAD
  solution (first
  row of Fig. \ref{Fig:sources_of_error_HYDRAD}) 
  shows that following an initial phase where
  the TR retains its pre-heating properties, the arrival of the
  coronal conduction front leads to a short interval (12--15s)
  when the downward heat flux ($F_{c,0}$)
  is balanced by the terms
  neglected in LareJ ($N$). 
  This interval is also associated with
  a small downward enthalpy flux 
  ($F_{c,0} \approx 50\,|F_{e,0}|$) that leads
  to enhanced radiative losses in the UTR. The components of 
  $N$ (right panel) show that in 
  the early phase, the base terms
  remain unimportant, and the $\ell\bar{E}/dt$
  term (change of UTR total
  thermal energy) dominates. Looking at 
  Fig. \ref{Fig:sources_of_error_LareJ}, we see that
  with LareJ the initial 12s is similar to HYDRAD, though the
  radiative losses decrease in the corona and hence our
  estimate of $\mathcal{R}_{utr}$ decreases, leading to a small
  upward
  enthalpy flux. Then, between 12--15s, LareJ shows a premature
  upflow with
  the enthalpy flux being driven immediately by the conductive
  losses, a direct consequence of the neglect of 
  $\ell\bar{E}/dt$ in
  the jump condition.
  \\
  \indent
  With HYDRAD, from 15s until the time of the first density
  peak (100s), both $N$ and radiative losses decline and become
  negligible after 50s, with the downward heat flux driving an
  upward enthalpy flux. However, it should be noted that up to
  40s, $N$ is still under an order of magnitude smaller than
  $F_{e,0}$
  so the retention of these terms in HYDRAD leads to a smaller
  upflow than in LareJ, as shown in 
  Fig. \ref{Fig:sources_of_error_LareJ}.
  At later times the
  HYDRAD values of $N$ are negligible. 
  \\
  \indent
  The apex and fp1 examples all show broadly similar behaviour:
  for fp1 heating the UTR is still driven by a downward heat
  flux. We have provided a summary of this general scenario in
  Table \ref{table:apex_fp1_simulations_jc_terms}, 
  which breaks down the atmospheric response to
  impulsive heating into 4 distinct phases that are listed by
  their time of importance: (i) the initial atmosphere is
  undisturbed, (ii) a short phase when the UTR internal energy
  changes are important, (iii) an evaporative phase with the
  components of $N$ being of diminishing importance leading to
  (iv) the first and subsequent density peaks. We also note 
  that
  for stronger (weaker) heating events the terms neglected in
  LareJ are more (less) important so the errors at the peak
  density are larger (smaller): see Table 1 in Paper I. 
  \\
  \indent
  However, within this framework, there are some subtle
  differences between short and long loops for apex and fp1
  heating. One is that in the long loop examples, the heat flux
  that hits the UTR is systematically smaller because the total
  energy released is chosen to be lower. Therefore, there is
  less reaction at the base to the incoming heat flux which in
  turn means that the neglect of $N$ in LareJ is less important
  and smaller errors in the peak density arise 
  (Table \ref{table:non_uniform_simulations}). This
  suggests that an important parameter is the 
  \lq thermal energy
  impulse\rq\ on the UTR, defined as 
  $\int F_{c,0} \, dt$.
  \\
  \indent
  The second minor (and related) exception arises for the Case
  3 apex example (short heating pulse in a long loop). In this
  case, the conductive travel time along the
  loop is longer
  than the half width duration of the heating pulse. From the
  second row of Fig. 
  \ref{Fig:sources_of_error_HYDRAD} \& 
  \ref{Fig:sources_of_error_LareJ}, as well as 
  Table \ref{table:fp2_simulations_jc_terms}, 
  we see that
  the LareJ upflow is no longer enhanced over that obtained by
  HYDRAD because the energy input into the UTR is not
  sustained. What is seen in 
  Fig. \ref{Fig:sources_of_error_HYDRAD}  
  is not in fact a decrease
  in the importance in $N$, but instead a deficit in the
  dominant
  terms from the jump condition. A similar argument also
  applies to the fp1 heating event for Case 3 but for the
  return time of the upward propagating conduction front.
  
%
%
\subsection{Sources of error: under-evaporation (fp2 heating)
  \label{section:Sources_of_error_under_evaporation}}
  \indent
  As already noted, footpoint heating at the base of
  the TR
  (fp2) is the most challenging case but a broad outline of the
  results is as follows. There is over-evaporation in a number
  of cases but we also now find cases with LareJ with (1)
  under-evaporation and (2) an underestimation of the coronal
  temperature. An important aspect of this is the
  update
  calculation of radiation within the UTR in LareJ 
  (see Appendix A, Paper I, and Sect. 
  \ref{section:Numerical method and experiments}) 
  which can lead to
  artificially high radiative losses in the UTR that are
  consistent with the coarse spatial resolution used (e.g.
  BC13, Paper I). The HYDRAD solutions enable us to quantify
  this error which is limited to only the fp2 heating
  examples.
  \\
  \indent
  So the difficulty with fp2 heating is that part of the energy
  released during the heating event ($\ell \bar{Q}$) 
  may be lost due to
  (artificially high) radiation in the UTR rather than
  transported to the corona by heat conduction, and the LareJ
  solutions indeed have a spurious reduction of the heat flux
  into the corona ($F_{c,0}$). 
  This also provides the explanation
  for the underestimation of the coronal temperature.
  Furthermore, it is clear from Eq. \eqref{eqn:1d_utr_jc}  
  that any reductions in
  $F_{c,0}$ may then also affect the upflow. 
  \\
  \indent
  For Case 3 fp2 heating event, the HYDRAD solution shows that
  for a short interval (0--10s) at the start of the heating
  period, the local energy released ($\ell \bar{Q}$)
  is balanced by $N$
  (with $\ell d\bar{E}/dt$ the dominant term as above). 
  In contrast,
  Fig. \ref{Fig:sources_of_error_LareJ}
  shows that the LareJ solution starts with a premature
  upflow (0--10s) that is powered by $\ell \bar{Q}$ 
  precisely because the
  $\ell d\bar{E}/dt$ term is neglected. At 10s an upflow begins
  in HYDRAD
  which remains present until 230s. In this evaporation phase,
  the upward enthalpy flux is first driven for a short time
  (10--60s) by the total local heating
  ($F_{c,0} + \ell \bar{Q}$) and then
  for a much longer period (60--200s) by the terms associated
  with $N$. 
  During this longer period, $\ell d\bar{E}/dt$ 
  first peaks as a
  negative term at 60s due to the rapid drop off in the energy
  release ($\ell \bar{Q}$). 
  $\ell d\bar{E}/dt$ then balances $F_{e,0}$ from 60--100s.
  After 100s the enthalpy flux at the base becomes important as
  the TR moves back upwards (following the end of the heating)
  and the base motions take over the driving of the upward
  enthalpy flux. On the other hand, the jump condition does not
  model this part of the upflow. Hence, the LareJ solution
  underestimates $F_{e,0}$ for the duration of this period
  (60--200s), as can be seen in Fig.
  \ref{Fig:sources_of_error_LareJ}.
  \\
  \indent
  For the Case 4 fp2 heating, the HYDRAD solution shows
  generally a similar response, however, in this slower heating
  event, the main evaporation phase takes place between
  50--600s, with the upward enthalpy flux driven by the total
  local heating ($F_{c,0} + \ell \bar{Q}$). 
  In contrast, these local heating
  mechanisms do not have equal weighting in driving the LareJ
  evaporation. Fig. \ref{Fig:sources_of_error_LareJ}
  shows that from 50--600s, a somewhat
  overestimated $\ell \bar{Q}$ dominates a significantly
  underestimated
  $F_{c,0}$. This behaviour arises as a direct consequence of
  the
  artificial energy losses in the UTR. The outcome is that
  LareJ
  underestimates the upward enthalpy flux between 200--600s. 
  \\
  \indent
  The quantities that control the evaporation in response to
  fp2 heating are predominantly the same for short and long
  loops. They are summarised together in Table
  \ref{table:fp2_simulations_jc_terms}
  which shows additional subtleties. 
  For example, LareJ 
  overestimates the initial upward enthalpy flux 
  in Cases 1 \& 2 for the reasons 
  discussed
  in Sect. \ref{section:Sources_of_error_over_evaporation}
  but
  underestimates
  $F_{e,0}$ 
  at later times due to the behaviour described above in this
  section
  for slow and fast heating. 
  The net effect is for a reduced
  over-evaporation in comparison to the 
  corresponding apex and fp1 heating
  events.

%
%
\section{Conclusions
  \label{section:Conclusions}} 
  \indent
  We introduced
  the jump condition approach for 1D hydrodynamic
  modelling in Paper I.
  This is a simple method that can be employed
  with an under-resolved TR to deal with the
  difficulty of obtaining the correct interaction 
  between a downward conductive flux and the 
  resulting upflow.
  Thus, the
  evaporative response to a coronal heating event
  can be modelled 
  \textit{without fully resolving the TR }(BC13).
  In this further
  analysis of the method, 
  the experiments presented were selected to
  be some of the most challenging cases.
  \\
  \indent
  In all of the experiments considered,
  the jump condition approach leads to coronal 
  densities that are comparable to
  fully resolved 1D 
  models (HYDRAD) but with computation 
  times that are between one and two  
  orders of magnitude faster. 
  Therefore, the applicability of the jump condition
  is not limited by introducing complexities,
  both spatially and temporally, in the energy
  release (heating).
  \\
  \indent
  On the other hand,
  the densities are predominantly higher  
  than those from a fully resolved 1D code (HYDRAD in this
  case)
  which is explained by the presence of an
  overestimated upward enthalpy flux. 
  This can be attributed to the neglect of terms corresponding
  to the rate of change of total energy in the unresolved
  atmosphere and mass motions at the base of the TR.
  It would certainly be advantageous to include these terms in
  the jump condition in order to remove the over-evaporation.
  However, if we could calculate these neglected terms
  accurately
  then it would not be necessary to implement such an approach.  
  Furthermore, the interaction between these terms is such that
  either both must be included or neither.
  Of course at some point diminishing returns will be reached
  and or the simplicity of the method weakened.
  \\
  \indent
  Despite the (relatively small) remaining error when
  comparing with a fully resolved 1D code (HYDRAD), 
  the implementation of the jump condition leads to a
  significant improvement compared with the equivalent (coarse
  resolution) simulations run without the jump condition. Both 
  the
  coronal density and temperature evolution are comparable
  with those obtained from fully resolved simulations,
  especially at the time of peak density and throughout the
  draining phase for both uniform (Paper I) and non-uniform
  heating (this paper).

%
%
\begin{acknowledgements} 
  The authors are grateful to Dr. Stephen Bradshaw for
  providing us with the HYDRAD code.
  We also thank the referee for their helpful 
  comments that improved the presentation.
  C.D.J. acknowledges the financial support of the Carnegie   
  Trust for the Universities of Scotland. 
  This project has received funding from the Science and 
  Technology Facilities Council (UK) through the consolidated 
  grant ST/N000609/1 and the European Research Council (ERC) 
  under the European Union's Horizon 2020 research and 
  innovation programme (grant agreement No 647214).
\end{acknowledgements}

%
%
\bibliographystyle{aa}
\bibliography{LareJPaper2}

\begin{thebibliography}{28}
\expandafter\ifx\csname natexlab\endcsname\relax\def\natexlab#1{#1}\fi

\bibitem[{{Antiochos} \& {Sturrock}(1978)}]{paper:Antiochos&Sturrock1978}
{Antiochos}, S.~K. \& {Sturrock}, P.~A. 1978, \apj, 220, 1137

\bibitem[{{Arber} {et~al.}(2001){Arber}, {Longbottom}, {Gerrard}, \&
  {Milne}}]{paper:Arber2001}
{Arber}, T.~D., {Longbottom}, A.~W., {Gerrard}, C.~L., \& {Milne}, A.~M. 2001,
  Journal of Computational Physics, 171, 151

\bibitem[{{Barnes} {et~al.}(2016){Barnes}, {Cargill}, \&
  {Bradshaw}}]{paper:Barnesetal2016}
{Barnes}, W.~T., {Cargill}, P.~J., \& {Bradshaw}, S.~J. 2016, \apj, 833, 217

\bibitem[{{Bourdin} {et~al.}(2013){Bourdin}, {Bingert}, \&
  {Peter}}]{paper:Bourdinetal2013}
{Bourdin}, P.-A., {Bingert}, S., \& {Peter}, H. 2013, \aap, 555, A123

\bibitem[{{Bradshaw} \& {Cargill}(2006)}]{paper:Bradshaw&Cargill2006}
{Bradshaw}, S.~J. \& {Cargill}, P.~J. 2006, \aap, 458, 987

\bibitem[{{Bradshaw} \& {Cargill}(2010a)}]{paper:Bradshaw&Cargill2010a}
{Bradshaw}, S.~J. \& {Cargill}, P.~J. 2010a, \apjl, 710, L39

\bibitem[{{Bradshaw} \& {Cargill}(2010b)}]{paper:Bradshaw&Cargill2010b}
{Bradshaw}, S.~J. \& {Cargill}, P.~J. 2010b, \apj, 717, 163

\bibitem[{{Bradshaw} \& {Cargill}(2013)}]{paper:Bradshaw&Cargill2013}
{Bradshaw}, S.~J. \& {Cargill}, P.~J. 2013, \apj, 770, 12

\bibitem[{{Bradshaw} \& {Mason}(2003)}]{paper:Bradshaw&Mason2003}
{Bradshaw}, S.~J. \& {Mason}, H.~E. 2003, \aap, 407, 1127

\bibitem[{{Bradshaw} \& {Viall}(2016)}]{paper:Bradshaw&Viall2016}
{Bradshaw}, S.~J. \& {Viall}, N.~M. 2016, \apj, 821, 63

\bibitem[{{Cargill} {et~al.}(2012a){Cargill}, {Bradshaw}, \&
  {Klimchuk}}]{paper:Cargilletal2012a}
{Cargill}, P.~J., {Bradshaw}, S.~J., \& {Klimchuk}, J.~A. 2012a, \apj, 752, 161

\bibitem[{{Cargill} {et~al.}(2012b){Cargill}, {Bradshaw}, \&
  {Klimchuk}}]{paper:Cargilletal2012b}
{Cargill}, P.~J., {Bradshaw}, S.~J., \& {Klimchuk}, J.~A. 2012b, \apj, 758, 5

\bibitem[{{Cargill} {et~al.}(2015){Cargill}, {Warren}, \&
  {Bradshaw}}]{paper:Cargilletal2015}
{Cargill}, P.~J., {Warren}, H.~P., \& {Bradshaw}, S.~J. 2015, Philosophical
  Transactions of the Royal Society of London Series A, 373, 20140260

\bibitem[{{Dahlburg} {et~al.}(2016){Dahlburg}, {Einaudi}, {Taylor},
  {Ugarte-Urra}, {Warren}, {Rappazzo}, \& {Velli}}]{paper:Dahlburgetal2016}
{Dahlburg}, R.~B., {Einaudi}, G., {Taylor}, B.~D., {et~al.} 2016, \apj, 817, 47

\bibitem[{{Hansteen} {et~al.}(2015){Hansteen}, {Guerreiro}, {De Pontieu}, \&
  {Carlsson}}]{paper:Hansteenetal2015}
{Hansteen}, V., {Guerreiro}, N., {De Pontieu}, B., \& {Carlsson}, M. 2015,
  \apj, 811, 106

\bibitem[{{Hood} {et~al.}(2016){Hood}, {Cargill}, {Browning}, \&
  {Tam}}]{paper:Hoodetal2016}
{Hood}, A.~W., {Cargill}, P.~J., {Browning}, P.~K., \& {Tam}, K.~V. 2016, \apj,
  817, 5

\bibitem[{{Johnston} {et~al.}(2017){Johnston}, {Hood}, {Cargill}, \& {De
  Moortel}}]{paper:Johnstonetal2017}
{Johnston}, C.~D., {Hood}, A.~W., {Cargill}, P.~J., \& {De Moortel}, I. 2017,
  \aap, 597, A81

\bibitem[{{Klimchuk} {et~al.}(2008){Klimchuk}, {Patsourakos}, \&
  {Cargill}}]{paper:Klimchuketal2008}
{Klimchuk}, J.~A., {Patsourakos}, S., \& {Cargill}, P.~J. 2008, \apj, 682, 1351

\bibitem[{{Lionello} {et~al.}(2009){Lionello}, {Linker}, \&
  {Miki{\'c}}}]{paper:Lionelloetal2009}
{Lionello}, R., {Linker}, J.~A., \& {Miki{\'c}}, Z. 2009, \apj, 690, 902

\bibitem[{{Miki{\'c}} {et~al.}(2013){Miki{\'c}}, {Lionello}, {Mok}, {Linker},
  \& {Winebarger}}]{paper:Mikicetal2013}
{Miki{\'c}}, Z., {Lionello}, R., {Mok}, Y., {Linker}, J.~A., \& {Winebarger},
  A.~R. 2013, \apj, 773, 94

\bibitem[{{O'Hara} \& {De Moortel}(2016)}]{paper:OHara&DeMoortel2016}
{O'Hara}, J.~P. \& {De Moortel}, I. 2016, \aap, 594, A67

\bibitem[{{Reale}(2014)}]{paper:Reale2014}
{Reale}, F. 2014, Living Reviews in Solar Physics, 11

\bibitem[{{Reale}(2016)}]{paper:Reale2016}
{Reale}, F. 2016, \apjl, 826, L20

\bibitem[{{Reale} {et~al.}(2016){Reale}, {Orlando}, {Guarrasi}, {Mignone},
  {Peres}, {Hood}, \& {Priest}}]{paper:Realeetal2016}
{Reale}, F., {Orlando}, S., {Guarrasi}, M., {et~al.} 2016, \apj, 830, 21

\bibitem[{{Reep} {et~al.}(2013{\natexlab{a}}){Reep}, {Bradshaw}, \&
  {Klimchuk}}]{paper:Reepetal2013a}
{Reep}, J.~W., {Bradshaw}, S.~J., \& {Klimchuk}, J.~A. 2013{\natexlab{a}},
  \apj, 764, 193

\bibitem[{{Reep} {et~al.}(2013{\natexlab{b}}){Reep}, {Bradshaw}, \&
  {McAteer}}]{paper:Reepetal2013b}
{Reep}, J.~W., {Bradshaw}, S.~J., \& {McAteer}, R.~T.~J. 2013{\natexlab{b}},
  \apj, 778, 76

\bibitem[{{Warren} {et~al.}(2011){Warren}, {Brooks}, \&
  {Winebarger}}]{paper:Warrenetal2011}
{Warren}, H.~P., {Brooks}, D.~H., \& {Winebarger}, A.~R. 2011, \apj, 734, 90

\bibitem[{{Warren} {et~al.}(2012){Warren}, {Winebarger}, \&
  {Brooks}}]{paper:Warrenetal2012}
{Warren}, H.~P., {Winebarger}, A.~R., \& {Brooks}, D.~H. 2012, \apj, 759, 141

\end{thebibliography}

\end{document}